%% file: main.tex
\documentclass[conference]{IEEEtran}
\IEEEoverridecommandlockouts

\usepackage{cite}
\usepackage{amsmath,amsfonts,amssymb,amsthm}
\usepackage{graphicx}
\usepackage{booktabs}
\usepackage{enumitem}
\usepackage{microtype}
\usepackage{textcomp}
\usepackage{xcolor}
\usepackage{xspace}
\usepackage{balance}
\usepackage{tikz}
\usetikzlibrary{arrows.meta,positioning,fit,backgrounds,calc}
\usepackage{url}
\usepackage[hidelinks]{hyperref}
\hypersetup{
  pdftitle={Mind the Hook: Source-Level Auditing of Privacy Defenses in Retrieval-Augmented Generation},
  pdfauthor={Yanhang Li, Zhichao Fan, Zexin Zhuang},
  pdfsubject={Source-level auditing of privacy defenses in retrieval-augmented generation},
  pdfkeywords={retrieval-augmented generation, privacy audit, source-level audit, differential privacy, membership inference}
}

\let\citep=\cite
\let\citet=\cite

\graphicspath{{build/figures/}}

\renewcommand{\paragraph}[1]{\par\vskip2pt\noindent\textbf{#1}\enspace\ignorespaces}

\newcommand{\nelstrict}{NEL\textsubscript{strict}\xspace}
\newcommand{\nelloose}{NEL\textsubscript{loose}\xspace}
\newcommand{\nodef}{\textsc{No-Defense}\xspace}
\newcommand{\dpretr}{\textsc{DP-R}\xspace}
\newcommand{\cadp}{\textsc{CA-DP}\xspace}
\newcommand{\privrag}{\textsc{Private-RAG}\xspace}
\newcommand{\pad}{\textsc{PAD}\xspace}
\newcommand{\lprag}{\textsc{LPRAG}\xspace}

\begin{document}

\title{Mind the Hook: Source-Level Auditing of\\
       Privacy Defenses in Retrieval-Augmented Generation}

\author{
  \IEEEauthorblockN{Yanhang Li}
  \IEEEauthorblockA{\textit{Northeastern University}\\
  Boston, MA, USA\\
  li.yanha@northeastern.edu}
  \and
  \IEEEauthorblockN{Zhichao Fan}
  \IEEEauthorblockA{\textit{University of Illinois Urbana-Champaign}\\
  Urbana, IL, USA\\
  zhichao8@illinois.edu}
  \and
  \IEEEauthorblockN{Zexin Zhuang}
  \IEEEauthorblockA{\textit{Southern Methodist University}\\
  Dallas, TX, USA\\
  zexinz@smu.edu}
}

\maketitle

\bstctlcite{IEEEtran:BSTcontrol}

\input{sections/abstract}

\begin{IEEEkeywords}
retrieval-augmented generation, privacy audit, source-level audit,
differential privacy, membership inference, named-entity leakage, canary
\end{IEEEkeywords}

\input{sections/intro}
\input{sections/related}
\input{sections/benchmark}
\input{sections/method}
\input{sections/results}
\input{sections/limitations}
\input{sections/conclusion}

\balance
\bibliographystyle{IEEEtran}
\bibliography{references}

\end{document}

%% file: sections/abstract.tex
\begin{abstract}
Black-box privacy scores for retrieval-augmented generation (RAG) are
difficult to interpret unless the audited defense's active pipeline
hook is known. We propose an \emph{active-path audit}: inventory
source-level hooks over retrieval, retrieved content, and generation;
map each metric to the leakage channel it observes; and validate
generated-text effects with exact-match canaries. In our benchmark
reimplementations, the DP-style defenses modify retrieval scores only:
their generation hooks are \texttt{TODO}-flagged stubs that return
responses unchanged. This active path explains why they affect
membership-inference behavior but track \nodef{} on generated-text
named-entity leakage, measured by \nelstrict. By contrast, the
end-to-end \lprag{} path is canary-validated on the email channel,
recovering 53/150 canaries under \nodef{} and 0/150 under \lprag.
These findings concern our reimplementations on our stack, not released
defenses or defense families; the contribution is a methodology and
case study, not a universal ranking.
\end{abstract}

%% file: sections/intro.tex
\section{Introduction}
\label{sec:intro}

A RAG privacy defense can be present in a codebase yet inactive on
the leakage channel being measured. If the implementation modifies
only retrieval, a privacy metric scored on the generated text cannot
see whether the defense works; if it modifies only the response, a
metric scored on the retrieval/index path cannot see it either. Most
current evaluations of RAG privacy defenses report only black-box
leakage numbers, leaving this kind of channel--implementation
mismatch invisible.

We make this concrete with a \emph{source-level active-path audit}
(\S\ref{sec:audit-method}): for each defense, inspect which pipeline
hook (retrieval scoring, retrieved-content rewriting, generation,
output post-processing) the implementation actually modifies; map
each privacy metric to the channel it observes; and interpret metric
movement only when the defense can act on that channel. End-to-end
effects on the generated-text channel are then validated with
exact-match canaries, which are not scored by the heuristic classifier.

We apply this audit to one fixed open-source RAG stack
(Phi-3-mini-4k-instruct \cite{abdin2024phi3} at 4-bit NF4, FAISS
\cite{johnson2019faiss} over \texttt{all-MiniLM-L6-v2}
sentence-transformer embeddings \cite{reimers2019sbert}, top-$k{=}5$, greedy decoding) with six
attack families (four extraction baselines, the agent attacker of
\citet{jiang2024ragthief}, and a black-box membership-inference
attack inspired by \citet{shokri2017membership}), six defense
implementations (a no-defense baseline, three DP-style implementations
we wrote---\dpretr, \cadp, \privrag---a regex masker \pad, and an
entity-substitution implementation \lprag\ inspired by
\citet{he2025lprag}), and three controlled corpora, arranged as a
$6 \times 6 \times 3 \times 2 \times 2 = 432$-cell grid
(\S\ref{sec:benchmark}; one run per cell, so all reported confidence
intervals are within-instantiation only). Three findings emerge that
would be hidden by a black-box-only evaluation:

\begin{itemize}[leftmargin=*,itemsep=0.2em,topsep=0.2em]
\item The three DP-style defense implementations in our benchmark
act \emph{only} on retrieval scores: their generation hooks are
\texttt{TODO}-flagged pass-throughs
(Table~\ref{tab:defense-hooks}). This explains why they reduce
MI AUC on Synthetic-Corp from $70.5$ to $41.5$--$59.3$
while tracking the undefended baseline on \nelstrict, which is
scored on generated text.
\item Regex masking (\pad) removes email-format strings on
Synthetic-Email but does not reduce person-name leakage in our
setup.
\item Entity substitution (\lprag) reduces the \texttt{EMAIL\_REAL}
sub-class of \nelstrict\ by $95.6\%$; an out-of-vocabulary email
canary (3 seeds, 50 trials each) recovers $53/150$ emails under
\nodef\ and $0/150$ under \lprag\ (Fisher $p < 10^{-14}$),
independently validating the end-to-end email effect. The
\texttt{PERSON\_NAME} sub-class is structurally entangled with
\lprag's substitution vocabulary and is exploratory.
\end{itemize}

These findings are a case study; the primary contribution is the
audit methodology. Without an active-path check, a black-box
evaluation would have read the DP-style implementations as
``MI-reducing privacy defenses,'' which is misleading: the same
implementations do not perturb generated text. The silent-stub finding
is not a claim that the cited source defense papers contain stubs; it is
that a benchmark wrapper can silently evaluate an implementation whose
active path does not overlap the metric it reports. The methodology is
general; the numeric findings are specific to this single open-source
stack.

\paragraph{Contributions.}
\textbf{(1) Active-path audit protocol.} A three-step procedure: hook
inventory, metric-to-channel map, and canary validation of
generated-text effects (\S\ref{sec:audit-method}).
\textbf{(2) Case study and silent-stub failure mode.} On a $432$-cell
grid, the protocol catches a silent-stub failure mode in three DP-style
implementations that a black-box benchmark would miss; Table~I
documents the active hooks.
\textbf{(3) Canary triangulation.} A targeted out-of-vocabulary
email canary ($3 \times 50$ trials) yields exact-match evidence
($53/150$ vs.\ $0/150$) for the one defense whose generation hook is
end-to-end, and explicitly does not certify the person-name
sub-class, which remains scorer-entangled. We do \emph{not} claim a
ranking of defense families or generalization beyond the audited
stack.

%% file: sections/related.tex
\section{Related Work}
\label{sec:related}

\textbf{Attacks on RAG and LM privacy.} Zeng et al.~\cite{zeng2024good}
document RAG leakage under adversarial prompting; Jiang
et al.~\cite{jiang2024ragthief} operationalize it with an agent-based
attacker; Carlini et al.~\cite{carlini2021extract} extract training data from
language models. We use Shokri et al.~\cite{shokri2017membership}'s black-box
membership-inference framing, contextualised by the auditing
perspective of Carlini et al.~\cite{carlini2022mifirstprinciples}, and
the canary methodology of~\cite{carlini2019secret} to validate
generated-text effects independently of a heuristic scorer.

\textbf{Privacy defenses for RAG.} Koga et al.~\cite{koga2024privacy}
study differentially private RAG (private-voting/partition mechanisms);
He et al.~\cite{he2025lprag} propose LPRAG, \emph{local-DP} entity
perturbation. Our \privrag\ is a DP-on-scores filter inspired by the
former, and our \lprag\ a deterministic entity-substitution
simplification of the latter (without its local-DP); \dpretr\ and
\cadp\ are simpler DP baselines we wrote. We audit our own
implementations, not the source papers.

\textbf{Algorithmic auditing and ML measurement.} Source-level
audits of ML systems---inspecting code and configurations rather
than only black-box behavior---are an established
algorithmic-auditing artefact in the sense of Raji
et al.~\cite{raji2020auditing}. A growing line of benchmark-reliability
audits makes related measurement concerns explicit---canary-based
memorization auditing after unlearning~\cite{li2026auditing},
configuration-conditional rank instability on alignment
benchmarks~\cite{li2026safetyrepro}, paired sample-size budgeting for
quantization benchmarks~\cite{zhuang2026preregistering}, and
economic-validity auditing of tabular foundation
models~\cite{wang2026auditing}. We adapt that posture to RAG privacy: an
active-path check is a small, mechanical audit that catches
silent-stub failure modes a single-metric benchmark would miss.

\textbf{Position.} We are not proposing a new attack, defense, or
metric. The contribution is the methodology---hook inventory,
metric-to-channel map, canary validation---and a case study on one
fixed open-source RAG stack.

%% file: sections/benchmark.tex
\section{Active-Path Audit Methodology}
\label{sec:audit-method}

Most RAG-privacy evaluations report a leakage number per
(defense, attack, corpus) cell and rank defenses by that number. The
number is only as meaningful as the assumption that the defense's
implementation actually intervenes on the channel the metric reads.
We make that assumption checkable with three steps.

\begin{enumerate}[leftmargin=1.4em,itemsep=2pt,topsep=2pt,parsep=0pt]
\item \textbf{Source-level hook inventory.} For each defense module,
inspect the implementation and classify which pipeline hook(s) it
modifies: \textbf{retrieval scoring} (noise on similarity scores or
top-$k$ truncation), \textbf{retrieved-content rewriting},
\textbf{generation} (token-level noise, paraphrasing, or
post-generation masking), or \textbf{none}. A hook whose body is
\texttt{TODO}-flagged or returns its input unchanged is recorded as
\emph{inactive}, yielding a hook--status table
(Table~\ref{tab:defense-hooks}).

\item \textbf{Metric-to-channel map.} For each metric, identify the
channel it observes. \nelstrict\ (\S\ref{sec:metric}) is computed on
\emph{generated text} and moves only if a defense modifies the
response; a black-box membership-inference AUC is designed to move when
the \emph{retrieval/index} path changes, though as a black-box score it
can also respond to output-surface nuisance (length, masking,
flattening), so we read it as channel-location evidence rather than a
calibrated membership-privacy estimate. A metric is interpretable on a
defense only if its active hooks overlap the metric's channel;
otherwise a shift is nuisance or a different mechanism.

\item \textbf{Canary validation.} A heuristic generated-text scorer
can be entangled with a defense's placeholder vocabulary, inflating
apparent protection. We validate generated-text effects with an
out-of-vocabulary canary (\S\ref{sec:c2}): inject unique strings
outside any defense's substitution dictionary, exact-match score the
response, and report recovery. The canary is not scored by the
heuristic classifier---the score path is bypassed---so it cannot be
inflated by scorer-vocabulary entanglement.
\end{enumerate}

\paragraph{Audit applied to our benchmark implementations.}
We applied this protocol to the six defense implementations in our
benchmark (Table~\ref{tab:defense-hooks}). Only three of the five
non-\nodef\ modules apply no explicit generation-stage transformation
globally: the three
DP-style modules (\dpretr, \cadp, \privrag) return the response via a
\texttt{\# TODO}-flagged pass-through (\dpretr's source literally reads
``For now, return unmodified response (placeholder)'', with analogous
paths in \cadp\ and \privrag). \pad\ is pattern-limited---its
generation hook fires only when its email-format regex matches---and
only \lprag\ performs end-to-end entity perturbation. Every
result in \S\ref{sec:results} therefore audits \emph{the specific
implementations we wrote}; upstream authors of cited defense papers
bear no responsibility for the \texttt{TODO} stubs our
implementations contain. The point is not that any source paper ships
stubs, but that a benchmark wrapper can silently evaluate an
implementation whose active path does not overlap the metric it
reports---which is exactly what an active-path check is meant to catch.
The channel split is what reconciles the
``DP reduces MI but not \nelstrict'' pattern as a property of the
audited stubs rather than of DP on RAG retrieval in general: MI's
black-box score is retrieval-sensitive (it shifts when DP noise on
retrieval changes which documents surface), whereas \nelstrict\ counts
named entities in the generated text, which carries no explicit
generation-stage perturbation when the generation hook is a stub (and
which we observe does not move appreciably; \S\ref{sec:channel-split}).

\begin{table}[t]
\centering
\small
\caption{Defense hook status. \emph{Stub} = a \texttt{TODO}
pass-through in \texttt{apply\_defense\_generation}, i.e.\ the generation
hook is inactive (no explicit output transformation; any change is
indirect, via altered retrieval context). \dpretr, \cadp, \privrag\
stubs are in our reimplementations, not upstream releases. The
\emph{retrieval-side} column covers both score noise and
retrieved-content rewriting; it is distinct from the generated-text
path read by \nelstrict.}
\label{tab:defense-hooks}
{\footnotesize\setlength{\tabcolsep}{3pt}\begin{tabular}{@{}lccl@{}}
\toprule
Defense & retrieval-side hook & generation hook & status \\
\midrule
\nodef   & --                 & --                  & baseline \\
\dpretr  & DP on scores       & \emph{stub}         & retr.-only \\
\cadp    & DP (per-turn)      & \emph{stub}         & retr.-only \\
\privrag & DP on scores       & \emph{stub}         & retr.-only \\
\pad     & --                 & regex (on match)    & pattern-lim. \\
\lprag   & entity sub.        & entity sub.         & end-to-end \\
\bottomrule
\end{tabular}}
\end{table}

\section{Artifact Under Audit}
\label{sec:benchmark}

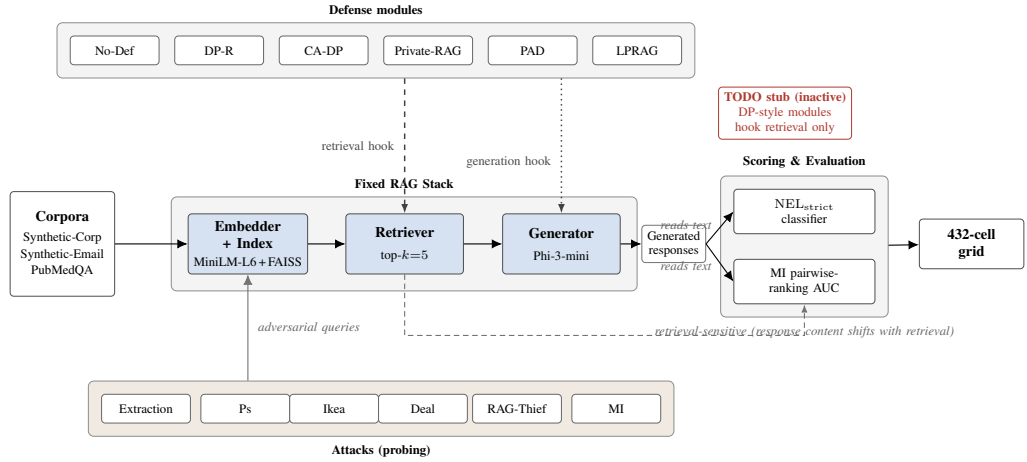
\begin{figure*}[t]
    \centering
    \input{sections/fig_pipeline}
    \caption{Audited RAG stack and the channel split. Defenses hook the
    retrieval and/or generation paths (dashed = retrieval, dotted =
    generation; active hooks per implementation in
    Table~\ref{tab:defense-hooks}); \nelstrict\ counts named entities in
    the generated text, while MI is a black-box score read on the same
    responses but \emph{retrieval-sensitive} (dashed bus): it shifts when
    retrieval perturbations change which documents surface, so each
    metric moves mainly when its channel's hook fires. The attacker reads
    responses, not internal retrieval scores. Hook status and the
    432-cell grid are detailed in \S\ref{sec:benchmark}.}
    \label{fig:pipeline}
\end{figure*}

We instantiate the audit on a fixed open-source-component RAG stack
(Figure~\ref{fig:pipeline}) whose entire cell grid shares one
language model, one retriever, one embedder, one top-$k$, and one
decoding policy. Holding the stack fixed lets us compare
implementations on equal terms; the cost is that we do not claim
cross-stack generalization. The benchmark is a \emph{case study} for
the audit method, not a universal ranking of defense families.

\subsection{Stack, attacks, defense implementations}
The language model is Phi-3-mini-4k-instruct \cite{abdin2024phi3} at
4-bit NF4 quantization; retrieval uses FAISS \cite{johnson2019faiss}
over \texttt{all-MiniLM-L6-v2} sentence-transformer embeddings
\cite{reimers2019sbert}; top-$k$ is fixed at 5 and generation is greedy. The raw driver produced 432
successful result JSONs (one per (attack, defense, corpus, mode,
$\varepsilon$) cell), plus one preserved \texttt{ERROR} stub from a
transient arithmetic-overflow on a single \lprag\ cell, recovered on
rerun; \S\ref{sec:results} aggregates the 432 successful cells.
The attack panel has six families: four extraction baselines
(\textsc{Ps}, \textsc{Pm}, \textsc{Ikea}, \textsc{Deal}), a
reproduction of the agent attacker of \citet{jiang2024ragthief}
(\textsc{RAG-Thief}), and a black-box membership-inference attack
(\textsc{Mi}; 10 member + 100 non-member docs per corpus, black-box
access to the RAG API, no shadow training, per-corpus pairwise-ranking
AUC) inspired by \citet{shokri2017membership} and contextualised by
the auditing perspective of \citet{carlini2022mifirstprinciples}.
The defense panel has six implementations, all in our benchmark
source tree:
\nodef; \dpretr\ (our implementation adding Laplace noise to
retrieval scores, $\varepsilon \in \{1,5\}$); \cadp\ (our
implementation of composition-aware DP budget across multi-turn
dialogue); \privrag\ (a DP-on-scores filter inspired by, not
reproducing, the DP-RAG of \citet{koga2024privacy}); \pad\ (our regex
masker for email and identifier patterns); and \lprag\ (a deterministic
entity-substitution simplification of the LPRAG of \citet{he2025lprag}
without its local-DP perturbation---fixed names
$\{\texttt{Alice}, \ldots\}$, placeholder domains
$\{\texttt{example.com}, \ldots\}$).
Results in \S\ref{sec:results} are therefore findings about
\emph{these benchmark implementations} on this stack, not about
upstream releases of the cited papers. $\varepsilon$ is inert for
the three non-DP defenses; under greedy decoding and a fixed seed
those $\varepsilon=1$ and $\varepsilon=5$ runs are
\emph{deterministic near-duplicates} rather than independent
replicates, and bootstraps that pool them collapse duplicates before
resampling (\S\ref{sec:metric}).

\paragraph{Threat model (compact summary).}
All five extraction attacks issue a fixed, bounded per-run query budget
and score the generated response for named-entity leakage; \textsc{Mi}
also runs black-box against the RAG API and scores the response, but
with a retrieval-sensitive membership heuristic (\S\ref{sec:metric})
rather than the \nelstrict\ entity count.

\subsection{Corpora}
\label{sec:corpora}

Three corpora of 50 documents each, treated as \emph{controlled
stimuli} varying in entity density and domain, not real-deployment
surrogates:
\begin{itemize}[leftmargin=1.2em,itemsep=1pt,topsep=2pt,parsep=0pt]
\item \textbf{Synthetic-Corp:} programmatic business corpus with
template-generated named entities; baseline for leakage behavior.
\item \textbf{Synthetic-Email:} 50 email-formatted documents whose
generator embeds Enron-era executive surnames (\texttt{skilling},
\texttt{fastow}, \ldots) into otherwise synthetic text; used for
canary validation (\S\ref{sec:c2}).
\item \textbf{PubMedQA}~\cite{jin2019pubmedqa}: 1{,}000 real abstracts,
first 50 indexed as the retrieval KB; a realistic QA workload for
probing whether DP hooks propagate into generated text.
\end{itemize}
\noindent\emph{Synthetic-Email provenance.} Synthetic-Email is
synthetic, produced by a Hugging Face fallback (a non-existent dataset
ID, then an absent local directory, then synthetic generation, as
reconstructed from run logs). It is \textbf{not} real Enron email, and
we make no claim of real corporate-email or real-PII extraction from
this corpus.

%% file: sections/fig_pipeline.tex
\definecolor{stackblue}{HTML}{D6E2F2}
\definecolor{groupgray}{HTML}{F4F4F4}
\definecolor{attackgray}{HTML}{F0EAE2}
\definecolor{notered}{HTML}{B03A2E}
\resizebox{0.74\textwidth}{!}{%
\begin{tikzpicture}[
  font=\footnotesize,
  io/.style={draw=black!70, thin, rounded corners=2pt, fill=white, align=center,
             minimum height=0.72cm, minimum width=2.0cm, inner sep=3pt},
  mod/.style={draw=black!70, thin, rounded corners=2pt, fill=stackblue, align=center,
             minimum height=1.15cm, minimum width=2.25cm, inner sep=3pt},
  metric/.style={draw=black!70, thin, rounded corners=2pt, fill=white, align=center,
             minimum height=0.85cm, minimum width=2.7cm, inner sep=3pt, font=\scriptsize},
  chan/.style={draw=black!55, thin, rounded corners=2pt, fill=white, align=center,
             minimum height=0.55cm, inner sep=3pt, font=\scriptsize},
  pill/.style={draw=black!60, thin, rounded corners=2pt, fill=white,
             inner sep=3pt, font=\scriptsize, minimum height=0.55cm, minimum width=1.7cm, align=center},
  flow/.style={-{Latex[length=2mm,width=1.8mm]}, semithick},
  hookR/.style={-{Latex[length=1.8mm]}, dashed, thick, black!75},
  hookG/.style={-{Latex[length=1.8mm]}, dotted, thick, black!75},
  sens/.style={-{Latex[length=1.6mm]}, densely dashed, semithick, black!55},
  tag/.style={font=\scriptsize\itshape, text=black!60},
  grp/.style={draw=black!45, thin, rounded corners=3pt, fill=groupgray},
]


\node[io, minimum height=2.0cm] (corpora) at (0,0)
   {\textbf{Corpora}\\[2pt]\scriptsize Synthetic-Corp\\\scriptsize Synthetic-Email\\\scriptsize PubMedQA};
\node[mod] (emb) at (3.55,0) {\textbf{Embedder}\\\textbf{+ Index}\\[1pt]\scriptsize MiniLM-L6\,{+}\,FAISS};
\node[mod] (ret) at (6.55,0) {\textbf{Retriever}\\[3pt]\scriptsize top-$k{=}5$};
\node[mod] (gen) at (9.55,0) {\textbf{Generator}\\[3pt]\scriptsize Phi-3-mini};
\node[chan] (resp) at (11.7,0) {Generated\\responses};
\node[metric] (mnel) at (14.2,0.62)  {$\mathrm{NEL_{strict}}$\\classifier};
\node[metric] (mmi)  at (14.2,-0.72) {MI pairwise-\\ranking AUC};
\node[io, minimum height=0.95cm] (grid) at (17.4,0) {\textbf{432-cell}\\\textbf{grid}};
\foreach \x/\name [count=\i] in
   {1.0/No-Def, 3.0/DP-R, 5.0/CA-DP, 7.0/Private-RAG, 9.0/PAD, 11.0/LPRAG}
  \node[pill] (d\i) at (\x,3.7) {\name};
\foreach \x/\name [count=\i] in
   {1.6/Extraction, 3.5/Ps, 5.2/Ikea, 6.9/Deal, 8.7/RAG-Thief, 10.6/MI}
  \node[pill] (a\i) at (\x,-3.15) {\name};

\draw[flow] (corpora) -- (emb);
\draw[flow] (emb) -- (ret);
\draw[flow] (ret) -- (gen);
\draw[flow] (gen) -- (resp);
\draw[flow] (resp.east) -- (mnel.west) node[pos=0.34, above left=-2pt, tag] {reads text};
\draw[flow] (resp.east) -- (mmi.west)  node[pos=0.34, below left=-2pt, tag] {reads text};

\begin{scope}[on background layer]
  \node[grp, fit=(emb)(ret)(gen), inner sep=9pt] (stack) {};
  \node[grp, fit=(mnel)(mmi), inner sep=7pt] (score) {};
  \node[grp, fit=(d1)(d6), inner sep=7pt] (defense) {};
  \node[grp, fit=(a1)(a6), fill=attackgray, inner sep=7pt] (attacks) {};
\end{scope}
\node[font=\scriptsize\bfseries, above=1pt of stack.north] {Fixed RAG Stack};
\node[font=\scriptsize\bfseries, above=1pt of score.north] {Scoring \& Evaluation};
\node[font=\scriptsize\bfseries, above=1pt of defense.north] {Defense modules};
\node[font=\scriptsize\bfseries, below=1pt of attacks.south] {Attacks (probing)};

\draw[flow] (score) -- (grid);

\draw[hookR] ($(defense.south-|ret.north)$) -- (ret.north)
   node[midway, left=2pt, font=\scriptsize, text=black!70] {retrieval hook};
\draw[hookG] ($(defense.south-|gen.north)$) -- (gen.north)
   node[pos=0.62, left=2pt, font=\scriptsize, text=black!70] {generation hook};
\node[draw=notered, thin, rounded corners=2pt, text=notered, align=center,
      font=\scriptsize, inner sep=3pt, anchor=west] at (12.55,2.5)
   {\textbf{TODO stub (inactive)}\\DP-style modules\\hook retrieval only};

\draw[sens] (ret.south) -- (6.55,-1.75) -- (14.2,-1.75) -- (mmi.south)
   node[pos=0.52, below, tag] {retrieval-sensitive (response content shifts with retrieval)};

\draw[flow, black!55] (attacks.north-|emb.south) -- (emb.south)
   node[midway, right=2pt, tag] {adversarial queries};

\end{tikzpicture}}%

%% file: sections/method.tex
\section{The \nelstrict\ Metric}
\label{sec:metric}

Prior RAG privacy papers report an \emph{extraction rate}: the fraction
of target secrets recovered. That is unambiguous with known targets and
exact-match scoring, but not when ``a named entity'' is the goal---a
scorer must decide whether \texttt{Jeffrey Skilling},
\texttt{Quantum Financial}, \texttt{Primary Care}, and
\texttt{user\_9266@demo.io} each count. The wrong classifier yields two
failure modes: \textbf{Type A (false protection)}, where a
placeholder-substituting defense looks protective only because the
scorer excludes its own placeholder vocabulary; and \textbf{Type B
(false leakage)}, where counting organizational/topical phrases
(\texttt{Enron Corporation}, \texttt{Social Security}) inflates apparent
leakage under \nodef.

\paragraph{Two-tier heuristic NEL.} We release a $\sim$50-line
rule-based regex+lexicon classifier (no second LM, whose behavior would
become part of the measurement) that buckets each extracted item into
one of eleven types. \textbf{\nelstrict} counts only
\texttt{EMAIL\_REAL} and \texttt{PERSON\_NAME} and is the headline
metric; it excludes \texttt{PERSON\_NAME\_AMBIGUOUS} and
\texttt{ORG\_TITLE} ($\sim$48\% of items), so it undercounts leakage.
\nelstrict\ is a \emph{heuristic score with limited validation}: audited
at 20/20 on an informal 10-per-class positive spot-check (Clopper--Pearson
one-sided 95\% lower bound on 10/10 $\approx0.74$), not a stratified
precision estimate and not a recall audit, hence a lower bound under a
stronger attacker. Only the email sub-class has external (non-classifier)
validation here, via the OOV canary (\S\ref{sec:c2}); other cross-defense
\nelstrict\ comparisons are measurements of the heuristic, not validated
leakage estimates.

\paragraph{Aggregation and uncertainty.} Each raw cell is one run; the
resampling unit is the run, and for each pool we bootstrap 2{,}000 times
and report the 95\% interval. Because each cell is a single seed, every
interval is \textbf{within-instantiation only}---variance across runs,
not over corpus realizations or attack-query samples---and is not a
population-level interval; we use overlap descriptively, never as a
significance test. The driver emits $\varepsilon{\in}\{1,5\}$ runs per
(attack, defense, corpus, mode); these are independent for the DP
defenses but \emph{deterministic near-duplicates} for \nodef, \pad, and
\lprag\ (no active DP call-site, greedy decoding, fixed seed). All
reported estimates use the \textbf{deduplicated} pool ($n{=}30$ per
non-DP defense, $n{=}60$ per DP defense); deduplication shifts no
headline \nelstrict\ mean by more than $0.1$ entities/run and flips no
comparison.

\paragraph{MI as a separate metric.} MI success is the attacker's
pairwise-ranking \textbf{AUC} ($\times100$: $50$ chance, $100$ perfect,
$<50$ a score reversal), over $10\times100=1000$ (member, non-member)
document-score pairs per corpus. \emph{Each document's score is a
black-box heuristic read on the RAG's generated responses} to
membership-probing queries (response length, target-keyword hits,
presence of numbers/emails/named entities); the attacker never reads
internal retrieval scores. The score is thus
retrieval-\emph{sensitive}---DP noise changes which documents surface in
the response---but, being a black-box output heuristic, also responds to
output-surface nuisance, which is why we read MI as
channel-\emph{location} evidence, not a calibrated membership-privacy
estimate (\S\ref{sec:limitations}). We do \emph{not} fold MI into
\nelstrict: the two have different units, budgets, and responses to
defenses.

%% file: sections/results.tex
\section{End-to-End Validation}
\label{sec:results}

All numbers use the deduplicated primary estimator (\S\ref{sec:metric}).
Table~\ref{tab:synopsis} is the audit dashboard; we then drill into the
load-bearing cells in evidentiary order: the canary
(\S\ref{sec:c2}), the MI-vs-\nelstrict\ channel split for the
retrieval-only DP-style implementations (\S\ref{sec:channel-split}),
and \pad's email-vs-name split (\S\ref{sec:c4}).

\begin{table*}[t]
\centering
\footnotesize
\caption{Audit dashboard. Email/name columns are the main \nelstrict\
split (per-run means for \nodef; signed reductions otherwise, positive
$=$ less leakage); MI AUC is Synthetic-Corp ($\times100$, $50=$chance).}
\label{tab:synopsis}
\begin{tabular}{@{}lllll@{}}
\toprule
Defense & Hook active       & Email NEL (heur.)              & Person-name NEL (heur.) & MI AUC ($50{=}$chance) \\
\midrule
\nodef       & ---                 & 2.25 per run (baseline)       & 1.75 per run (baseline) & 70.5 \\
\dpretr      & retrieval DP only   & $+8.1\%$ (Ex)                 & $-9.5\%$ (Ex)           & 59.3 \\
\cadp        & retrieval DP only   & $+5.9\%$ (Ex)                 & $-17.1\%$ (Ex)          & 41.5$^{\dagger}$ \\
\privrag     & retrieval DP only   & $+11.9\%$ (Ex)                & $-24.8\%$ (Ex)          & 49.7$^{\dagger}$ \\
\pad         & regex (gen, emails) & $+100.0\%$ on Syn-Email patt. & $-42.9\%$ (Ex)          & 30.0$^{\dagger}$ \\
\lprag       & entity sub.\ (both) & $\mathbf{+95.6\%}$ canary-val. & n/a (SE)                & 8.2$^{\dagger}$ \\
\bottomrule
\end{tabular}

\vspace{2pt}
{\footnotesize $^{\dagger}$ values $<50$ are score reversals (members
ranked below non-members), not ranking-able privacy wins. \emph{Ex} =
exploratory (heuristic scorer, not independently validated); \emph{SE}
= scorer-entangled with \lprag's vocabulary; \emph{canary-val.} =
validated by the OOV canary (Table~\ref{tab:canary}). Hook source in
Table~\ref{tab:defense-hooks}.}
\end{table*}

\subsection{Canary validation: the email channel of \lprag\ holds on out-of-vocabulary strings}
\label{sec:c2}

The cleanest evidence is an exact-match out-of-vocabulary (OOV) email
canary: it is scored on the generated text with no heuristic classifier
in the loop, and its domain is outside any defense's substitution
dictionary, so a defense cannot ``win'' it by reshaping a placeholder
vocabulary.

\paragraph{Protocol.} Following Carlini et al.~\cite{carlini2019secret},
we inject OOV email canaries of the form
\texttt{operator\_N@acmetest-internal.canary} (50 canaries, one per
document) into a 50-document synthetic KB and attack with \textsc{Ps}
at three seeds, using a 50-query \emph{targeted stress-test} pool (the
first 10 match the shared extraction pool; the other 40 are
email/identifier-targeted prompts that do not leak the canary prefix or
domain). This stress-tests exploitability, not a deployment rate: the
$35.3\%$ \nodef\ recovery is what a targeted adversary extracts.
Scoring is exact string match.

\begin{table}[tb]
\centering
\footnotesize
\caption{OOV canary exact-match recovery (3 seeds, 50 canaries/seed;
targeted stress-test pool, not a deployment rate).}
\label{tab:canary}
{\footnotesize\setlength{\tabcolsep}{5pt}\input{build/latex_tables/canary.tex}}
\end{table}

\paragraph{Result.} Across 3 seeds with 50 canaries each ($n=150$),
\nodef\ recovers 53/150 (35.3\%) and \lprag\ recovers 0/150. A Fisher
one-sided test treating the canary opportunities as independent gives
$p<10^{-14}$; because they are grouped by seed, the load-bearing
evidence is the consistent per-seed pattern (Table~\ref{tab:canary}),
not the asymptotic $p$-value. Since each canary's domain is outside
\lprag's placeholder dictionary, \lprag's reduction on the
\texttt{EMAIL\_REAL} sub-class of \nelstrict\ is not scorer-vocabulary
bias but the substitution path applying to the tested OOV
email-shaped strings.

\paragraph{Aggregate \nelstrict\ split.} On the aggregate, \lprag\
cuts the \texttt{EMAIL\_REAL} mean from 2.25 to 0.10 per run ($95.6\%$,
canary-validated above). Its \texttt{PERSON\_NAME} mean drops from
1.75 to 0.00 ($100\%$), but this is definitional, not validated: the
classifier's \texttt{PLACEHOLDER\_NAME} set \emph{is} \lprag's
\texttt{NAME\_VOCABULARY}, so every name \lprag\ writes is excluded
from \texttt{PERSON\_NAME} by construction. We do \emph{not} claim
coverage of person-name leakage under \lprag; an analogous OOV
person-name canary is future work.

\subsection{Channel split: retrieval-only DP-style implementations lower MI AUC but not \nelstrict}
\label{sec:channel-split}

Our three DP-style implementations (\dpretr, \cadp, \privrag) ship with
\texttt{TODO}-stubbed \texttt{apply\_defense\_generation} hooks and so
apply no explicit generation-stage transformation
(Table~\ref{tab:defense-hooks}). The active-path map predicts, and we
observe, a clean separation between the retrieval-sensitive black-box
response score (MI) and the named-entity count on the same generated
text (\nelstrict): only the MI heuristic moves appreciably under these
retrieval-only perturbations.

\paragraph{\nelstrict\ does not move appreciably.} Averaging over
attacks, corpora, modes, and (for DP) $\varepsilon$, \nodef\ emits
3.93 \nelstrict\ items/run (95\% within-instantiation interval
$[2.50,5.37]$). The three DP implementations track this baseline
within $\pm7\%$ (\dpretr\ $-1.3\%$, \cadp\ and \privrag\ $-6.1\%$;
negative $=$ slightly more leakage), every interval overlapping
\nodef. We read this descriptively, not as a significance test.

\paragraph{MI AUC moves.} On Synthetic-Corp, \nodef\ MI AUC is $70.5$
$[68.8,71.7]$; the DP implementations move it to $41.5$--$59.3$,
including reversals below $50$, consistent with DP noise shrinking
$|\text{AUC}-50|$. This is not in tension with the \nelstrict\ null:
MI's black-box response score is retrieval-sensitive (it shifts when
the retrieval hook perturbs which documents surface), whereas
\nelstrict\ counts named entities in that same text, which the stubs
do not transform. Because MI is black-box its AUC can also move under
output-surface nuisance, so we treat it as channel-location evidence
(\S\ref{sec:limitations}), not a calibrated privacy estimate. (On
PubMedQA, MI is at the $100$ ceiling under \nodef\ and $\geq97$ under
DP---an artifact of an index-based oracle over a known 50-abstract KB,
not a privacy finding.)

\subsection{\pad\ blocks emails on Synthetic-Email, not person-name leakage}
\label{sec:c4}

On Synthetic-Email, \nodef\ emits 5.30 \texttt{EMAIL\_REAL} and 4.00
\texttt{PERSON\_NAME} items/run. \pad\ drives \texttt{EMAIL\_REAL} to
0.00 but its \texttt{PERSON\_NAME} count sits at 6.15 ($+2.15$ vs
\nodef), so aggregate \nelstrict\ only drops 9.30$\to$6.15: the regex
masker blocks the email family it targets but is not a general
person-entity defense. We flag the $+2.15$ name delta as descriptive
(our precision spot-check is only $20/20$ on a 10-per-class sample, so
we do not claim it is mechanistic). A practitioner reading only the
aggregate would miss this two-sided pattern. More broadly, the audit
makes two metric pathologies legible: \textbf{scorer-vocabulary
entanglement} (a classifier excluding a defense's substitution
dictionary reports near-perfect protection by construction---hence the
\texttt{PERSON\_NAME} caveat above) and \textbf{ambiguous-class
inflation} (our scorer routes $\sim$48\% of items into ambiguous
org/person classes, so reporting \nelloose\ rather than the strict
pair would shift apparent baselines by tens of percent without
changing behavior). We therefore treat \nelstrict\ as a
precision-oriented lower bound, not a leakage estimate.

%% file: build/latex_tables/canary.tex
\begin{tabular}{lcccc}
\toprule
Defense & seed 101 & seed 102 & seed 103 & Mean \\
\midrule
\nodef & 21/50 & 16/50 & 16/50 & 17.7/50 (35.3\%) \\
\lprag & 0/50 & 0/50 & 0/50 & 0.0/50 (0.0\%) \\
\bottomrule
\end{tabular}

%% file: sections/limitations.tex
\section{Scope and Threats to Validity}
\label{sec:limitations}

The audit methodology is general; the numeric findings are case-study
evidence on one stack, not a universal ranking of RAG privacy defenses.

\paragraph{Single seed, one stack, synthetic corpora.} Each of the 432
raw cells is one run, so every bootstrap interval is
within-instantiation only: it captures variance across pooled runs but
not over alternative synthetic-corpus realizations or attack-query
samples, and is not a population-level interval (the canary
sub-experiment of \S\ref{sec:c2} is the only one with three
independent seeds). The stack is fixed (one LM, embedder, top-$k$); a
larger model might respond differently to DP-retrieval noise. The five
extraction attacks are bounded fixed-query baselines; stronger adaptive
attackers would plausibly raise \nelstrict. Two of three corpora are
synthetic, and the ``Enron''-labeled runs are \emph{synthetic} emails
from a generator that hard-codes Enron-era surnames
(\S\ref{sec:corpora}). Cross-stack and real-corpus generalization is
out of scope.

\paragraph{Implementations we audited, not defense families.} The
channel-split findings are measurement reports about our three
benchmark implementations of DP-class defenses, not claims about
DP-on-RAG theory or upstream source papers: the three DP-class modules
return the response unmodified and \pad's generation path fires only on
a regex match (Table~\ref{tab:defense-hooks}). A fully-wired DP-RAG
baseline requires completing those \texttt{TODO}s, which we mark as
follow-on work. The active-path methodology does not depend on this
gap---the same protocol would flag any implementation with an inactive
hook on a measured channel.

\paragraph{\nelstrict\ is precision-only; MI is not identified.} Our
scorer audit is a spot-check (20/20 on a 10-per-class positive sample),
not a stratified precision estimate, and we did not audit recall, so
\nelstrict\ is a lower bound under a stronger attacker; a larger
stratified audit and the person-name OOV canary are follow-on work. The
MI score correlates with, but is not unique to, membership---defenses
that shorten or mask output could lower MI AUC via surface-corruption
nuisance---so we phrase the channel split narrowly as ``reduced success
on this MI protocol'' and read MI as channel-location evidence only.

%% file: sections/conclusion.tex
\section{Conclusion}
\label{sec:conclusion}

We presented an active-path audit for RAG privacy defenses---hook
inventory, metric-to-channel map, canary validation---and applied it
to one fixed open-source RAG stack with six defense implementations
on a 432-cell grid. Three takeaways:

\begin{itemize}[leftmargin=*,itemsep=0.2em,topsep=0.2em]
\item \textbf{The audit catches a silent-stub failure mode.} Three of
six \emph{audited benchmark implementations} modify only retrieval; a
black-box benchmark would report them as ``MI-reducing defenses''
($70.5\!\to\!41.5$--$59.3$) without registering that \nelstrict\ on
generated text does not move. The failure mode is benchmark-wrapper
active-path drift, not a property of DP defenses as such.

\item \textbf{Heuristic scorers need canary triangulation.} \lprag's
email-channel reduction is validated by an OOV canary ($53/150$ vs.\
$0/150$, $p<10^{-14}$), but its person-name reduction is
scorer-entangled and reported as exploratory, not certified.

\item \textbf{Regex masking is channel-specific.} \pad\ blocks
email-format strings but not person-name leakage; the audit makes the
specificity legible.
\end{itemize}

\noindent Before asking whether a RAG privacy defense works, ask where
it acts. The NEL classifier, analysis scripts, canary generator, and
$432$ cleaned runs will be released at de-anonymization.